\documentclass{article}
\usepackage{frascatiphys}
\begin{document}
\title{ 
The pomeron conjecture and two gluon glueballs
}
\author{
Pedro Bicudo\\
{\em Dep. F\'{\i}sica and CFIF, Instituto Superior T\'ecnico,
Av. Rovisco Pais, 1049-001 Lisboa, Portugal} 
}
\maketitle
\baselineskip=11.6pt
\begin{abstract}
In this talk 
the pomeron conjecture is reviewed and 
constituent gluon models are derived.
In a simple two-gluon glueball spectrum
the pomeron trajectory and the daughter
trajectories are computed. The open
problems of two-gluon glueballs are 
discussed, including transversality and 
Yang's theorem, the spin tensor interactions,
the structure of the string and decays.
The related systems of charmed hybrids and 
of the gluelump are also addressed.
To conclude, different aspects of glueballs
that could be measured at PANDA are highlighted.
\end{abstract}
\baselineskip=14pt
\section{Introduction}

\par
The pomeron trajectory produces a precise prediction for the
location of several glueball masses, see Table \ref{masstab}.
The equation for the pomeron trajectory in the $ J,\, t=M^2$ space is
\cite{Donnachie,Pelaez},
\begin{eqnarray}
J= \alpha_p(t),
\\
\alpha_p(t) = 1.08+0.25 t \ .
\end{eqnarray}
The intercept $\alpha_p(t)$ is of the order of 1, and this explains 
the high energy hadronic cross sections. The pomeron is also expected 
to correspond to a series of glueball masses. The slope of 0.25 is 
compatible with the Casimir scaling, where the string tension is 
proportional to the Casimir invariant $ \lambda_i \cdot \lambda_j$.

\begin{table}[t]
\centering
\caption{ \it Spectrum of glueballs extracted from the Pomeron trajectory in a Regge plot.
}
\vskip 0.1 in
\begin{tabular}{|l|cccccc|} \hline
 J P C   & 2++  & 4++  & 6++  & 8++  & 10++ & 12++ \\
 M[GeV]  & 1.92 & 3.41 & 4.53 & 5.26 & 5.97 & 6.61 \\
\hline
\end{tabular}
\label{masstab}
\end{table}

\par
For instance the $\bar p -p$ total cross sections increase monotonously at 
high energies. The cross sections are fitted by,
\begin{equation}
\sigma=21.70 \, s^{0.0808} + 98.39 \, s^{-9.4525}
\end{equation}

\par
\begin{figure}[t]
 \vspace{9.0cm}
\includegraphics{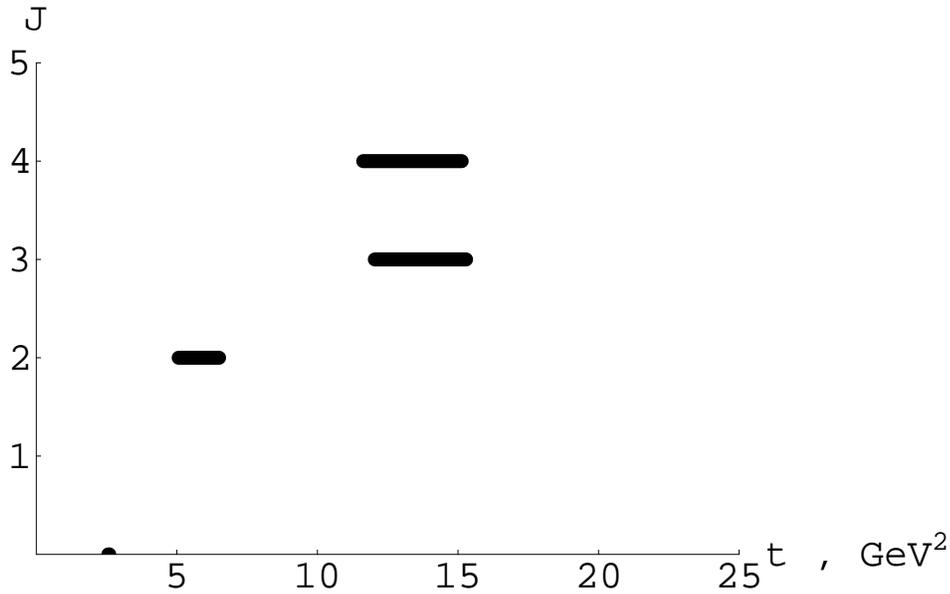}
\caption{\it
      Lattice glueballs with PC=++ displayed in a Regge plot .
    \label{lattice} }
\end{figure}

The lattice results
\cite{Morningstar}
and model results
\cite{Llanes}
can also be included in a Regge Plot, see fig.\ref{lattice}. 
They are not inconsistent with the Pomeron
\cite{lattice Regge}, 
and with a daughter trajectory.

\par
In what concerns model calculations, they depend on the gluon mass and on the 
gluon-gluon interaction. Although one would expect colour gauge invariance 
to force the gluon to be massless, the generation of a finite mass for the gluon 
is nevertheless possible.
For instance in superconductors the Meissner effect generates a mass for the 
photon
\cite{Gennes}. 
In the QCD case, we have indications from the lattice 
\cite{Bernard}
and from the solution
of the truncated Schwinger Dyson equation 
\cite{Alkofer}
that the gluon has a constituent mass
of 0.7 to 0.8 GeV.

\par
A flux tube picture is also possible, where the string ends in a pair of spin 1, colour 
octet massive gluons. There are indications, both from lattice simulations
\cite{Casimir}
and from truncated 
Coulomb Gauge QCD 
\cite{Llanes}
that the string tension is proportional to the Casimir invariant 
$\lambda_1 \cdot \lambda_2$. Then $\sigma_{glueball}={9 \over 4} \sigma_{meson}$. 

\par
It is therefore plausible to construct constituent gluon models, similar to the constituent
quark model, where the lightest glueballs have just two gluons. In the self-consistent models
the gluon mass is generated in the models, and in other models the mass is just assumed.
This results in bound-states similar to mesons, where the quarks are replaced by heavier 
gluons and where the string tension is also larger.

\par
Essentially I will focus on two approximations of QCD,
\\ - the ITEP approach, of a quantum mechanical glueball string model by
Kaidalov, Kalashnikova, Nefediev, Shevchenko, Simonov
\cite{ITEP}
\\ - the approach started in the NCSU, of a self-consistent glueball in Coulomb Gauge by
Cotanch, Llanes-Estrada, Swanson, Szczepaniak
\cite{Llanes,NCSU,Szczepaniak}

\section{A simple 2-gluon glueball spectrum}

\par
\begin{figure}[t]
 \vspace{9.0cm}
\includegraphics{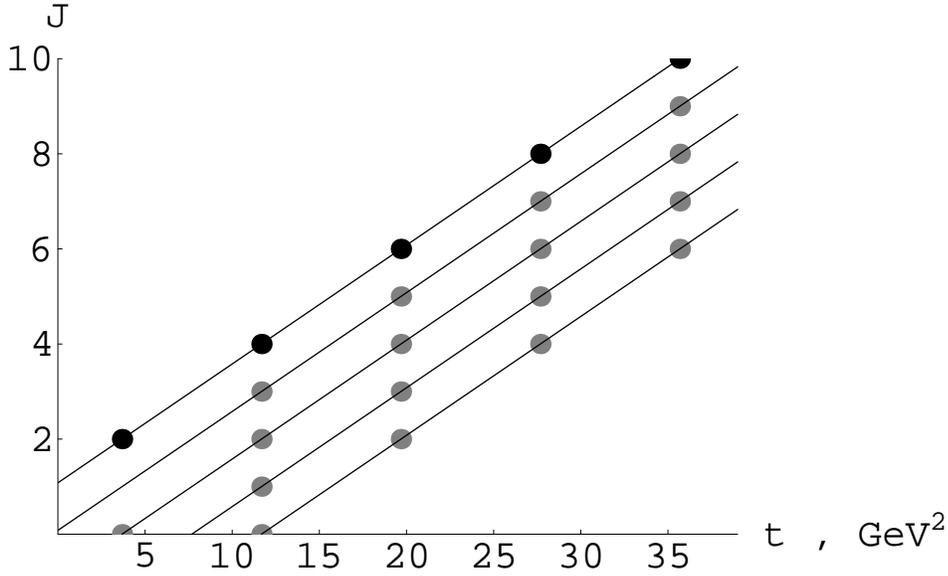}
 \caption{\it
      Pomeron trajectory and daughters in a Regge plot.
    \label{daughters} }
\end{figure}

From the pomeron a quite simple and precise picture emerges,
see fig.\ref{daughters}. I think that any glueball study, experimental, 
theoretical or on the lattice should compare with the pomeron.

\par
Let us assume, in the constituent gluon model perspective, that we have at least 2 gluons
with the highest possible J,
\begin{eqnarray}
S&=&1+1=2 \ ,
\nonumber \\ 
J&=&L+2 \ ,
\end{eqnarray}
and with L even for P=+.
Then these states should be aligned in the pomeron trajectory. The corresponding
spectrum is detailed in Table \ref{masstab}.
This constitutes the simplest and less speculative prediction for glueballs. 

\par
To verify experimentally that we have a straight line, at least 3 points should be 
measured. The first 3 points are expected to be observed  if  $M = 5$ GeV is reached.  
However, we also know the example of the meson Regge trajectory, where 
the first point is below the line, and then we need more points to see the line.  
If an experiment is able to observe glueballs with $M = 5.3 \pm 0.3$ GeV, 
it may be possible to see 4 points.  The error appears because the slope of the 
pomeron trajectory is is not precise. For instance Szczepaniak and Swanson 
\cite{Szczepaniak}
finds a smaller slope than the one first
estimated for the pomeron. 

\par
Moreover, in the constituent 2-gluon picture the pomeron trajectory
\begin{equation}
S=2, J=L+2 ,
\end{equation}
is accompanied by 5 daughter trajectories,
\begin{eqnarray}
&& S=2, J=L+2,
\nonumber \\
&& S=2, J=L+1,
\nonumber \\
&& S=2, J=L,
\nonumber \\
&& S=2, J=L-1,
\nonumber \\
&& S=2, J=L-2,
\nonumber \\
&& S=0, J=L.
\end{eqnarray}
The pomeron and daughters are depicted in fig.\ref{daughters},
in the case where the spin-dependent interactions are neglected.

\section{Open Problem in two-gluon glueballs}

\subsection{Transversality}

\par
The transversality of the gluon propagator implies that $JLS^{PC}=101++$ is forbidden,
according to Yang ' s theorem
\cite{Llanes}.
In the lattice, and in the Coulomb Gauge model, a light J=1 glueball is not present.
In quantum mechanical studies, this state is not avoided. 
The discovery of a J=1++ glueball would rule out the transversality of constituent gluons.

\subsection{Spin-Tensor Interactions}

\par
The spin-tensor interactions also depend of the model. The exact position of the daughters 
will measure the spin-dependent interactions.

\par
The $\vec S_1 \cdot \vec S_2$ interaction splits the S=2 daughter 
( $\langle  \vec S_1 \cdot \vec S_2  \rangle =1$ ) from the S=0 
daughter ( $ \langle \vec S1 \cdot \vec S_2 \rangle =-2 $). 
In particular it produces a 0++ glueball lighter than the 2++. 

\par
The  $\vec S \cdot \vec L$ interaction increases with L and thus changes the slope of the 
interaction. It also acts differently on the different daughters,
\begin{eqnarray}
j=l+2 &:&  	\langle \vec S \cdot \vec L \rangle  = 2l \ ,
\nonumber \\
j=l+1 &:& 	 \langle \vec S \cdot \vec L \rangle  = 2l- 4 \ ,
\nonumber \\ 
j=l &:&	 \langle \vec S \cdot \vec L \rangle  =-6  \ or \ 
\langle \vec S \cdot \vec L \rangle  =0 \ ,
\nonumber \\
\cdots
\end{eqnarray}
In particular the $\vec S \cdot \vec L$ interaction produces non-linear trajectories.

\par
The tensor interaction also produces a non-linear pomeron trajectory, moreover it couples 
$l$ with $l+2$ and $l-2$.

\par
In the meson spectrum, the $\vec S_1 \cdot \vec S_2$ interaction seems to be the most relevant 
spin-dependent interaction. In the lattice simulations and in the quantum mechanical 
studies of glueballs, the $\vec S_1 \cdot \vec S_2$ also seems to be the relevant one. 
In the truncated Coulomb gauge approach this is not the case.
The detection of the glueballs on the daughter trajectories would clarify the nature of
the gluon-gluon interaction.

\begin{figure}[t]
 \vspace{9.0cm}
\includegraphics{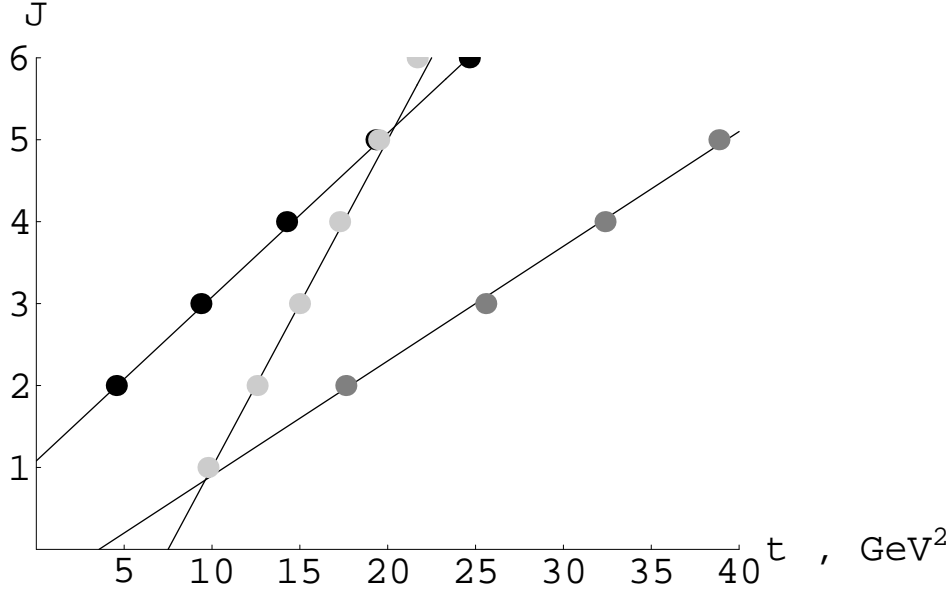}
 \caption{\it
      Leading Regge plot trajectories for the glueballs (in black circles), 
the gluelumps (in dark gray circles), and the charmonia (in light gray circles).
    \label{plus gluelump} }
\end{figure}

\subsection{The string}

\par
Are the gluons connected by 1 octet string or by 2 triplet ones?
In type II superconductors the double string is favoured
\cite{Gennes}, 
and this results in the
doubling of the string constant. In the lattice, there is evidence for 
a single octet-octet string with a strength proportional to $ \lambda_i \cdot \lambda_j$.
Although these two approaches only differ by 12.5 \%, a precise determination of the
slope of the gluonic Regge trajectories would also clarify the nature of
the gluon-gluon interaction.

\subsection{Decays}

The decays of Glueballs decay are not fully understood yet. In the litterature, the scalar glueball 
is essentially the only one with predicted decays
\cite{decays}. 
Nevertheless one can estimate that the glueballs will have a larger width than conventional hadrons, 
because they follow more decay mechanisms. In the double-string model, from string breaking an 
enhancement of the width by a factor of 2x2 is expected. Any of the two constituent gluons (attached 
at the end of the string) may also decay in a quark-antiquark pair. Because the decays of glueballs 
remain an open problem, measuring the decay widths and the decay processes of glueballs would 
be extremely interesting. 

\section{Charmed hybrids and gluelumps}

\par
The lowest hybrid states are difficult to separate from the standard chamonium spectrum.
However the excited states where the gluon is far from the diquark show different properties.
The gluelump is the heavy quark limit of the charmed or bottomed hybrid. 
This subject has been receiving an increased interest in the litterature
\cite{gluelump}.
In this case the heavy quark-antiquark pair forms a nearly point-like and massive colour octet,
equivalent to a very massive gluon. The reduced mass of the real gluon is close to the $c$ 
quark mass in charmonium. However the string tension is larger than the charmonium one. 
In a simple constituent gluon picture this results in different 
trajectories from the glueball trajectories and from the charmonium trajectories.
Neglecting the spin-tensor interactions, one gets the pc=++ states depicted
in fig. \ref{plus gluelump}.

\section{Conclusion}

\par
Identifying the glueballs with the largest possible angular momentum, up to $5.5\pm 0.3$ GeV 
will test the Pomeron conjecture. 

\par
The decay widths are expected to be large, and the decays should produce a large number 
of pions. Nevertheless the states are well separated. The decays of excited glueballs 
remain a very interesting open problem. The light quark hybrids constitute intermediate 
decay channels.

\par
The study of glueballs with lower angular momentum (daughter trajectories) will test 
many aspects of QCD, and will fix the spin dependence of the gluon-gluon interaction.

\par
Moreover there are odd parity trajectories, and more massive trajectories 
with 3 gluons (contributing to the odderon).

\par
In the charmed hybrid sector, and in the limit of the Gluelump, the study of larger 
angular momentum may also exhibit a linear Regge behavior.  

\par
The decays of the excited glueballs may result in the production of several pions, with a 
large total J.
\\ Question: Is it possible to identify the initial J of the glueball?
\\ Question: Is it possible to reconstruct the initial shape of the string, say with with 
Bose-Einstein correlation?

\section{Acknowledgements}
I thank Paola Gianotti and the PANDA collaboration for motivating this talk. I also thank discussions 
on constituent gluon masses and on constituent gluon models with Alexei Nefediev, Pedro Sacramento, 
Steve Cotanch and Felipe Llanes-Estrada.

\end{document}